# Modifications of the Dielectric Properties of Biological Membranes by Heating

S. Blanca Savescu
School of Mathematics, Kingston University London, Penrhyn Road, Kingston upon Thames, KT1 2EE, UK

## Introduction

Biological cell suspensions are known to show dielectric dispersions due to the Maxwell-Wagner mechanism. Many examples are summarized in a number of papers by Schwan [7, 9, 10]. By the application of an appropriate analysis to the dielectric dispersion, it is possible to estimate electrical phase parameters related to protoplasm and cell membrane. A dielectric theory of interfacial polarization for a suspension of conducting particles (protoplasm) covered with poorly conducting shells (plasma membrane) was developed by Pauly and Schwan [7], and was satisfactory applied for low volume fractions of suspended particles ( less than about 0.2).

The purpose of this paper is to examine the change in yeast plasma membrane permittivity after heating treatment, by using the Pauly and Schwan's theory.

## Materials and Method

Yeast cells (Saccharomyces cerevisiae) were grown in liquid cultures at 27°C and harvested in the stationary phase after two days. The collected cells were washed with distilled water, then incubated for one day in 10 mM KCl solution. Dead cells were prepared, before the measurements, by incubating for 10 minutes in water bath at 70°C. Both specimens (intact and heated cells) were rapidly washed with distilled water and resuspended, for measurements, in 10 mM KCl solutions.

The mean diameter of the cells was determined aver about 200 cells by measuring under an optical microscope. The mean diameter without wall was found to be 3.5 µm for dead cells. The values of the cell wall thickness (0.22 µm for intact cells and 0.25 µm for treated cells) were taken from literature [5].

Measurements of electrical admittance, over 0.1 – 10 MHz, were carried out with a Hewlett Packard vector-impedance meter, model 4194 A.

**Results and Discussion**

The permittivity and conductivity of viable yeast cell suspensions show a remarkable dependence on frequency (Fig. 1). Such disperse behaviour is due to the Maxwell-Wagner mechanism. Further we consider that the cell wall has no influence on the dielectric dispersion, In the present experimental conditions.

Under the assumptions that the shell phase is thin and poorly conducting, compared with the inner and outer phases, Pauly and Schwan [6] derived the following equations:

$$\delta\varepsilon = \frac{9}{4\varepsilon_0} \cdot \frac{p \cdot R \cdot C_m}{\left[1 + R \cdot G_m \left(\frac{1}{\sigma_l} + \frac{1}{2\sigma_a}\right)\right]^2} \qquad (1)$$

$$f_C = \frac{1}{2 \cdot \pi \cdot R \cdot C_m} \cdot \left[\left(\frac{1}{\sigma_l} + \frac{1}{2\sigma_a}\right)^{-1} + R \cdot G_m\right] \qquad (2)$$

$$\sigma_l = \sigma_a \left[1 - \frac{3}{2} \cdot p \cdot \frac{1 + R \cdot G_m \left(\frac{1}{\sigma_i} - \frac{1}{\sigma_a}\right)}{1 + R \cdot G_m \left(\frac{1}{\sigma_l} + \frac{1}{2\sigma_a}\right)}\right] \qquad (3)$$

where R is the cell radius, p volume fraction of cells, $\sigma_l$ suspension conductivity at law frequency, $\delta\varepsilon$ - the dielectric increment, $\sigma_i$ and $\sigma_a$ inner and outer conductivity, $C_m$ and $G_m$ membrane capacity and conductance and $\varepsilon_0$ = 8.854 · $10^{-12}$ F/m.

In the most biological situations, $G_m$ may be neglected and equations (1), (3), (2) become:

$$\delta\varepsilon = \frac{9}{4\varepsilon_0} \cdot p \cdot R \cdot C_m \qquad (4)$$

$$\sigma_l = \sigma_a \left(1 - \frac{3}{2} \cdot p\right) \qquad (5)$$

$$f_C = \frac{1}{2 \cdot \pi \cdot R \cdot C_m} \cdot \left(\frac{1}{\sigma_l} + \frac{1}{2\sigma_a}\right)^{-1} \qquad (6)$$

In order to estimate the electrical phase parameters of viable and treated yeast cells, we use the equations

(4)-(6). The experimental data points are fitted with the Cole-Cole formula [7]:

$$\varepsilon^* = \frac{\delta\varepsilon}{1+\left(j\cdot\frac{f}{f_C}\right)^\alpha} + \varepsilon_h \qquad (7)$$

where $\delta\varepsilon$ is the permittivity increment, $f_C$ the characteristic frequency, $\alpha$ - the Cole-Cole parameter, $\varepsilon_h$ permittivity at high frequencies and $j = \sqrt{-1}$. The phenomenological parameters are shown in Table 1.

Table 1. Dielectric parameters for viable and heat treated yeast cells at ionic equilibrium between cytoplasm and outer medium.

|  | $\delta\varepsilon$ | $\sigma_l$ (mS/m) | $\sigma_a$ (mS/m) | $f_C$ (MHz) |
|---|---|---|---|---|
| Living cells | 1043 | 67 | 100 | 0.7 |
| Dead cells | 262 | 87 | 110 | 0.85 |

As shown in Figure 1, the heating treatment gives rise to a reduction of the dielectric dispersion. Assuming that the equation (4) is applicable to both living and dead cells and considering that the capacity of dead cells is the same as that of viable cells [5], we may write:

$$\frac{\delta\varepsilon_v}{\delta\varepsilon_t} = \left(\frac{R_v}{R_t}\right)^4 \qquad (8)$$

where v and t refer to the viable and heated cells, respectively. The equation (8) is defined under the assumption that the number of cells is the same for both specimens so that:

$$\frac{p_v}{p_t} = \left(\frac{R_v}{R_t}\right)^3 \qquad (9)$$

From equation (8), the value of $\frac{\delta\varepsilon_v}{\delta\varepsilon_t}$ is expected to be about 1.27, whereas, as seen in Table 1, the dielectric increments experimentally obtained provide a ratio of 4. This discrepancy seems to be due to the increase of the membrane conductance, caused by heating treatment. Indeed, by analyzing eq. (1), an increase in the membrane conductance $G_m$ gives a reduction of the dielectric increment.

Figure1 Dielectric constant (a) and conductivity (b) vs. frequency for viable and heat treated yeast cells.

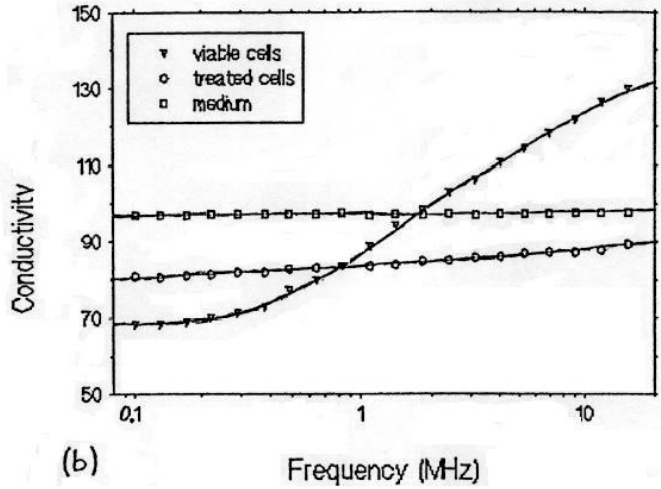

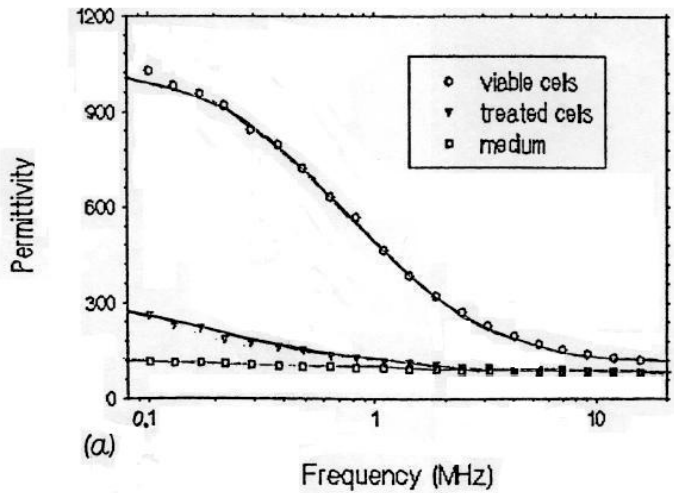

Combination of equations (1) and (3) gives the following formula for the membrane conductance:

$$G_m = 2 \cdot \left[ \sigma_1 - \sigma_a \left( 1 - \frac{3}{2} p \right) \right] \left( \frac{C_m}{p \cdot R \cdot \varepsilon_0 \cdot \delta\varepsilon} \right)^{\frac{1}{2}} \qquad (10)$$

Using $C_m = 1.1$ µF/m$^2$ [4], p = 0.18 (from eq. 9) and the values of conductivities at ionic equilibrium from Table 1, the membrane conductance of heated cells, $G_m$, is estimated to be $1.8 \cdot 10^4$ S/m. For membrane thickness of about 8 nm [5], we obtain a membrane conductivity of $1.4 \cdot 10^{-4}$ S/m. This value is of about 500 times greater than that of the viable cells, as reported by Ying Huang et al [5].

These features suggest that the heating treatment makes some damage on the membrane, so that the permitivity increases. At the same time, the membrane capacity, which is determined by the bulk of the membrane materials, does not change after heating treatment [4,5]. Therefore, we can conclude that the processes that affect the permittivity of the membrane occur only at specific sites, which occupy a minor part of the total membrane area. A similar effect has been observed for treatments with $Ag^+$ [1] and ionic detergent [2].

**Conclusions**

The dielectric spectroscopy offers an appropriate method for the study of biological membranes and their structural modifications which appear after treatments with physical and chemical agents.

It was proved that the changes in the curves of permittivity and conductivity of the heated yeast cell suspension are due to the increase in the conductance of the cell membrane. Hence, by thermal treatment, the biological membrane may undergo a alteration of the membrane proteins, although a phase transition of the lipids is also possible.

**References:**

1. Arnold W.M., Geier B.M, Wendt B. and Zimmermann U. (1986) *Biochim. Biophys. Acta*, **889**, pp.35
2. Asami, K. (1977) *Bull.Inst. Chem. Res.*, Kyoto Univ., **55**, 3, pp.283
3. Asami K. (1979) *Bull.Inst. Chem. Res.*, Kyoto Univ., **55**, 4, pp.297
4. Holzel, R., Lamprecht I. (1992) 1104, pp.195
5. Huang, Y., Holzel, R. Pethig R., Wang, X. (1992), *Phys. Med. Biol.* **37**, 7, pp.1499
6. Schwan, H.P. (1957) in *Advances in Biological and Medical Physics,* Vol. V J.H. Lawrence and C.A Tobias. Ed., Academic Press, New York, pp. 147

7. Pauly H., Schwan, H.P. (1959) *Z. Naturforsch.*, 14b, pp125
8. Schwan, H.P. (1963) in *Physical Techniques in Biological Research*, Vol. VI, Part B, W.L. Nastuk Ed., pp.323
9. Schwan, H.P. (1985) *IEEE Trans. Electrical Ins., Vol.* EI-20, 6, pp.913
10. Schwan, H.P. (1988) *Ferroelectrics*, Vol. 86, pp.205